\documentclass{ws-mpla}
\usepackage[super]{cite}
\usepackage{graphicx}
\begin{document}

\markboth{XIAO-MING XU, H. J. WEBER}
{INELASTIC MESON-MESON SCATTERING IN HADRONIC MATTER}

\catchline{}{}{}{}{}

\title{INELASTIC MESON-MESON SCATTERING\\
IN HADRONIC MATTER}

\author{\footnotesize XIAO-MING XU}

\address{Department of Physics, Shanghai University, 99 Shangda Road,\\
Baoshan, Shanghai 200444, 
China\\
xmxu@mail.shu.edu.cn}

\author{H. J. WEBER}

\address{Department of Physics, University of Virginia, Mccormick Road,\\
Charlottesville, VA 22904, USA}

\maketitle

\pub{Received (Day Month Year)}{Revised (Day Month Year)}

\begin{abstract}
We review studies of inelastic meson-meson scattering. In nonperturbative
schemes with chiral-perturbation-theory Lagrangians
and in models based on effective meson Lagrangians, 
inelastic meson-meson scattering leads to the successful identification of 
resonances in meson-meson reactions, adequate inclusion of final
state interactions in particle decays, and so on. For mesons of which each
consists of a quark and an antiquark, inelastic meson-meson scattering may
be caused by quark-antiquark annihilation, quark-antiquark creation, quark
interchange, and quark-antiquark annihilation and creation. In 
transition amplitudes for meson-meson scattering mesonic 
quark-antiquark relative-motion wave functions depend on hadronic matter,
and transition potentials are given in perturbative quantum chromodynamics. 
Via transition amplitudes the cross sections for inelastic meson-meson
scattering depend on the temperature of hadronic matter. Some prominent 
temperature dependence of the cross sections has been found. Inelastic 
meson-meson scattering becomes even more significant in proton-proton 
collisions and lead-lead collisions at the Large Hadron Collider.

\end{abstract}

\ccode{PACS Nos.: 13.75.Lb; 13.85.Hd; 12.39.Pn; 12.39.Fe}

\section{Introduction}	

Meson-meson scattering is an important tool in studying resonances. Resonances
are produced by the strong interaction of the colliding mesons. This amounts to
two-to-one meson-meson scattering, for example, $\pi\pi \to \rho$ and
$\pi K \to K^\ast$.

When the center-of-mass energy of two pions increases from the sum of
two pion masses, elastic pion-pion scattering takes place. When the energy
exceeds the sum of two kaon masses (four pion masses), 
a $K\bar K$ pair (two more pions) may be produced. 
The scattering process becomes elastic $\pi\pi$ scattering, 
elastic $K\bar K$ scattering, or elastic four-pion scattering. The connection
between elastic $\pi\pi$ scattering and elastic $K\bar K$ scattering is the
inelastic scattering $\pi\pi \leftrightarrow K\bar K$. A couped-channel 
formalism is the appropriate framework for describing this meson-meson 
scattering. Inelastic two-to-two meson-meson scattering is important in 
studying meson-meson scattering \cite{OO1997,DGL,GLMOR}, photon-photon 
reactions\cite{GLMOR,OO1998,DV,DDV}, $J/\psi$ decays \cite{MO,GPSCZ,SLTO}, etc.

Recently, the LHCb Collaboration at the Large Hadron Collider (LHC) has 
measured $pp$ 
collisions to examine the symmetry under the combined charge-conjugation and
parity-transformation operations \cite{LHCb1,LHCb2,LHCb3,LHCb4} (CP) in the 
decays 
$B^{\pm} \to \pi^\pm \pi^+ \pi^-$, $B^{\pm} \to \pi^\pm K^+ K^-$, 
$B^{\pm} \to K^\pm \pi^+ \pi^-$, and $B^{\pm} \to K^\pm K^+ K^-$. The
CP asymmetry observed in the decay modes is unexpectedly large in
some kinematical regions of the final mesons. The large CP violation has been
attributed to weak interactions, resonances of final
mesons, and $\pi\pi \leftrightarrow K\bar K$ \cite{NBCFL}.

Quark-gluon plasmas are created in Au-Au collisions at the Relativistic Heavy
Ion Collider and in Pb-Pb collisions at the LHC. The 
plasmas eventually transform into hadronic matter at the critical temperature. 
Hadronic matter changes until kinetic freeze-out occurs. The change of 
hadronic matter is governed by hadron-hadron scattering. All possible 
hadron-hadron scattering occurs in hadronic matter. Many experiments have been
done for 
nucleon-nucleon collisions, $p\bar p$ collisions, and meson-nucleon collisions
\cite{KMR,LXW}. However, the role these collisions play in hadronic matter 
is still largely unknown.
Experimental data on inelastic meson-meson scattering are scarce. Therefore,
many hadron-hadron collisions in hadronic matter and even in vacuum have to be
studied theoretically. This is a huge task for theorists!

Mesons are the dominant species in hadronic matter. The usual scattering is 
two-to-one meson-meson scattering and two-to-two meson-meson 
scattering. In lead-lead collisions and in xenon-xenon collisions at the LHC,
the meson momentum measured by the ATLAS Collaboration, the CMS Collaboration,
and the ALICE Collaboration goes up to 1000 GeV/$c$ 
\cite{ATLAS1000,CMS1000,ALICE1000}. A meson of such large momenta 
in collision with another meson in hadronic
matter may yield three or more mesons. Two-to-three meson-meson scattering
affects the chemical equilibrium of hadronic matter, and causes energy loss 
of the large-momentum meson. Therefore, we need to study two-to-three
meson-meson scattering in hadronic matter as well.

We review inelastic meson-meson scattering in vacuum and in hadronic matter. 
Inelastic meson-meson scattering in the coupled-channel unitary
approaches and the inverse amplitude method are discussed in Sec. 2. Cross
sections from effective meson Lagrangians are introduced in Sec. 3.
We talk about meson-meson reactions in Secs. 4-9 for the following processes:
quark-antiquark annihilation, quark-antiquark creation, quark-antiquark 
annihilation and creation, and quark interchange. A summary is given 
in the last section.

\section{Inelastic Meson-Meson Scattering in Coupled Channels}	

Describing mesons as quark-antiquark currents, the most general effective
Lagrangian that respects Lorentz invarinace, parity symmetry, and chiral
symmetry was put forward by Gasser and Leutwyler in 1984 \cite{GL1}. This is
the Lagrangian of $SU(2)$ chiral perturbation theory ($\chi$PT). 
Including the strange quark, SU(3) $\chi$PT was established in Ref. \cite{GL2}.
Furthermore, including the $\eta^\prime$ meson, $U(3)$ $\chi$PT was established
in Refs. \cite{HLPT,KL}. The $\chi$PT Lagrangians are given in powers of 
external momenta $p$ and quark masses. The Lagrangian to next-to-leading order
in $SU(3)$ $\chi$PT is
\begin{equation}
{\cal L}={\cal L}_2+{\cal L}_4 .
\end{equation}
The term ${\cal L}_2$ is of order $p^2$, and is given by
\begin{equation}
{\cal L}_2=\frac{f_0^2}{4}<\partial_\mu U^+ \partial^\mu U + M(U+U^+)> ,
\end{equation}
where $f_0$ is the pion decay constant, $M$ is the quark-mass-dependent 
matrix \cite{OO1997}, and the angular brackets stand for the
trace of the quantity enclosed. The matrix $U=\exp (i\sqrt{2}\Phi/f_0)$
displays the pseudo Goldstone boson fields by
\begin{equation}
\Phi (x) = \begin{pmatrix}
\frac{1}{\sqrt 2}\pi^0+\frac{1}{\sqrt 6}\eta & \pi^+ & K^+ \\
\pi^- & -\frac{1}{\sqrt 2}\pi^0+\frac{1}{\sqrt 6}\eta & K^0 \\
K^- & \bar{K}^0 & -\frac{2}{\sqrt 6}\eta 
\end{pmatrix} ,
\end{equation}
where $x$ are the space-time coordinates. The term ${\cal L}_4$ is of order 
$p^4$:
\begin{eqnarray}
{\cal L}_4 & = & L_1 <\partial_\mu U^+ \partial^\mu U>^2
+L_2 <\partial_\mu U^+ \partial_\nu U><\partial^\mu U^+ \partial^\nu U> 
   \nonumber  \\
& + & L_3 <\partial_\mu U^+ \partial^\mu U \partial_\nu U^+ \partial^\nu U>
+L_4 <\partial_\mu U^+ \partial^\mu U><U^+ M +M^+ U> \nonumber  \\
& + & L_5 <\partial_\mu U^+ \partial^\mu U (U^+ M +M^+ U)>
+L_6 <U^+ M +M^+ U>^2 \nonumber  \\
& + & L_7 <U^+ M -M^+ U>^2 +L_8 <M^+ U M^+ U +U^+ M U^+ M> ,
\end{eqnarray}
where $L_h$ $(h=1, \cdot \cdot \cdot, 8)$ are constants.

Usually two or three channels are considered in solving problems 
\cite{OO1997,DGL,GLMOR,DV,DDV,GPSCZ,SLTO,AOR,TDKJO}. For example, the 
two-channel case is $\pi\pi \to \pi\pi$, $K\bar{K} \to K\bar{K}$, and $\pi\pi 
\leftrightarrow K\bar{K}$, and the three-channel case is $\pi\eta \to \pi\eta$,
$K\bar{K} \to K\bar{K}$, $\pi\eta^\prime \to \pi\eta^\prime$,
$\pi\eta \leftrightarrow K\bar{K}$, $\pi\eta \leftrightarrow \pi\eta^\prime$,
and $K\bar{K} \leftrightarrow \pi\eta^\prime$. 
Denote by $N$ the number of channels 
involved. Elastic meson-meson scattering is indicated
by $A_i+B_i \to A_i+B_i$ with $i=1, \cdot \cdot \cdot, N$. In the 
coupled-channel unitary approaches one needs to calculate potentials from
${\cal L}_2$:
\begin{equation}
V_{ji}=<A_j B_j \mid {\cal L}_2 \mid A_i B_i>.
\end{equation}
From the Lippmann-Schwinger equations the $T$-matrix elements for 
$A_i+B_i \to A_j+B_j$ with $j=1, \cdot \cdot \cdot, N$ satisfy
\begin{equation}
T_{ji}=V_{ji}+\sum_{k=1}^NV_{jk}G_{kk}T_{ki},
\end{equation}
where $k=1, \cdot \cdot \cdot, N$, and
$G_{kk}$ are two-meson Green functions. $T_{ji}$ with $j=i$
($j \neq i$) is the $T$-matrix element for elastic (inelastic) meson-meson
scattering. Poles of the $T$-matrix
elements are identified as resonances \cite{OO1997}. The $T$-matrix elements
yield phase shifts for elastic scattering and 
inelasticity for inelastic scattering \cite{WI}.
To generate all the resonances with isospin 0 and masses below 2 GeV,
$S$-wave meson-meson scattering for total isospin $I=0$ and 1/2 of the two 
mesons is studied in Ref. \cite{AO} with 13 coupled channels.

A multi-channel generalization of the inverse amplitude method was given in
Refs. \cite{OOP1,OOP2} where the $T$-matrix is a $N \times N$ matrix.
Due to the unitarity the $T$-matrix is related to its inverse.
Deriving the inverse from scattering amplitudes obtained in $\chi$PT, 
the $T$-matrix elements acquire nonperturbative aspects which can produce
resonances, elastic phase shifts, and inelasticity of inelastic meson-meson
scattering.

In some decays and reactions where mesons are final states, final state
interactions become appreciable and even important 
\cite{Barnes2002,Barnes2003}. To facilitate studies of these decays and 
reactions, experimental data on some meson-meson scattering amplitudes are 
parametrized \cite{PR,SBGKL,DEW}. For example, a study of $\pi \pi \to K\bar K$
scattering by means of partial-wave dispersion relations of the Roy-Steiner
type is performed in Ref. \cite{PR}, and precise parametrizations of 
the $S$, $P$, and $D$ partial waves in the $\pi\pi \to K\bar K$ scattering 
amplitude are obtained from the data.

\section{Inelastic Meson-Meson Scattering from Effective Meson Lagrangians}

Irrespective of chiral symmetry, effective meson Lagrangians with explicit
resonances have been constructed to study meson-meson scattering. Let 
$\phi_{\rm s}$, $\phi_{\rm p}$, $\phi_{\rm v}^{\mu}$, and 
$\phi_{\rm t}^{\mu \nu}$ stand for the fields of the scalar meson, 
the pseudoscalar meson, the vector meson, and the tensor meson, respectively. 
For scattering between two pseudoscalar mesons interaction terms of 
the Lagrangian given in Ref. \cite{LDHS} are
\begin{equation}
{\cal L}_{\rm pps}={\rm g}_{\rm pps}m_{\rm p}\phi_{\rm p}(x)\phi_{\rm p}(x)
\phi_{\rm s}(x) ,
\end{equation}
\begin{equation}
{\cal L}_{\rm ppv}={\rm g}_{\rm ppv}\phi_{\rm p}(x)
\partial_\mu\phi_{\rm p}(x)\phi_{\rm v}^\mu(x) ,
\end{equation}
\begin{equation}
{\cal L}_{\rm ppt}={\rm g}_{\rm ppt}\frac{2}{m_\pi}\partial_\mu\phi_{\rm p}(x)
\partial_\nu\phi_{\rm p}(x)\phi_{\rm t}^{\mu\nu}(x) ,
\end{equation}
where $m_{\rm p}$ ($m_\pi$) is the mass of the pseudoscalar meson (pion);
${\rm g}_{\rm pps}$ (${\rm g}_{\rm ppv}$, ${\rm g}_{\rm ppt}$) is the coupling
constant of the two pseudoscalar mesons and the scalar meson (vector
meson, tensor meson).
In scattering of two pseudoscalar mesons scalar mesons, vector mesons, and 
tensor mesons are propagators in the $s$ channel, the $t$ channel, and the
$u$ channel. 
The $T$-matrix elements in the coupled-channel case are obtained from the 
Lippmann-Schwinger equations, and are used to get cross sections for 
meson-meson reactions like $\pi^+ \pi^- \to K\bar K$ . With 
${\cal L}_{\rm ppv}$ isospin-averaged cross sections for $\pi \pi \to K 
\bar K$, $\rho \rho \to K \bar K$, $\pi \rho \to K \bar {K}^*$, and 
$\pi \rho \to K^* \bar {K}$ in vacuum have been obtained in Ref. \cite{BKWX}
via exchange of either a kaon or a vector kaon between the two initial 
mesons. The cross sections depend on masses of initial and final mesons.
The baryon-rich medium created in intermediate-energy nucleus-nucleus 
collisions modifies meson masses. Cross sections for the four reactions in 
medium may be obtained by using in-medium meson masses.

Starting from a meson Lagrangian different from that in Ref. \cite{LDHS}, 
$\pi K^* \to \rho K$, $\pi K^* \to \omega K$, $\pi K^* \to \phi K$, 
$\rho K^* \to \pi K$, $\omega K^* \to \pi K$, and $\phi K^* \to \pi K$
have been studied in Ref. \cite{TKANN}. In calculations of cross sections
for these reactions in vacuum
the exchange of $K_1(1270)$ and $K_2^*(1430)$ in the 
$s$ channel, $h_1(1170)$ in the $t$ channel, etc. is taken into account.

\section{Meson-Meson Reactions due to Quark Interchange}

Meson $A$ is made of quark $q_1$ and antiquark $\bar{q}_1$, which may have 
different flavors, and meson $B$ consists of quark $q_2$ and antiquark 
$\bar{q}_2$, which may also have different flavors.
In the collision of mesons $A$ and $B$ quark $q_1$ and quark $q_2$ interchange,
and then $q_2$ and $\bar{q}_1$ combine to form meson $C$ and $q_1$ and 
$\bar{q}_2$ 
combine to form meson $D$. This is the quark interchange mechanism 
\cite{BS1992} which leads to the scattering $A(q_1\bar{q}_1)+B(q_2\bar{q}_2) 
\to C(q_2\bar{q}_1)+D(q_1\bar{q}_2)$. When an interaction takes place between
constituents of mesons $A$ and $B$ prior to quark interchange, the scattering
is designated as the prior form. When an interaction occurs between 
constituents of $q_2\bar{q}_1$ and $q_1\bar{q}_2$ 
after quark interchange, the scattering is
designated as the post form. In the prior form of the scattering
the color interaction between a constituent 
of meson $A$ and a constituent of meson $B$ turns the color-singlet states 
$A$ and $B$ into color-octet states of $q_1\bar{q}_1$ and $q_2\bar{q}_2$, 
respectively. During propagation of quarks and antiquarks interchange
of $q_1$ and $q_2$ causes 
$q_2$ ($q_1$) to find $\bar{q}_1$ ($\bar{q}_2$) to get the color-singlet state
$C$ ($D$). In the post form quark
interchange produces a color-octet state of $q_2\bar{q}_1$ and another
color-octet state of $q_1\bar{q}_2$. The color interaction between a 
constituent of $q_2\bar{q}_1$ and a constituent of $q_1\bar{q}_2$
makes $q_2\bar{q}_1$ ($q_1\bar{q}_2$) colorless so that the bound state 
$C$ ($D$) is formed.

The $S$-matrix element is \cite{LX}
\begin{equation}
S_{\rm fi} = \delta_{\rm fi} -\frac{(2\pi)^4 i 
\delta (E_{\rm f} - E_{\rm i}) \delta^3 (\vec{P}_{\rm f} - \vec{P}_{\rm i})}
{V^2\sqrt{2E_A2E_B2E_C2E_D}}
\left( {\cal M}^{\rm prior}_{q_1\bar{q}_2}
+{\cal M}^{\rm prior}_{\bar{q}_1 q_2}
+{\cal M}^{\rm prior}_{q_1q_2}+{\cal M}^{\rm prior}_{\bar{q}_1\bar{q}_2}
\right) ,
\end{equation}
for scattering in the prior form or
\begin{equation}
S_{\rm fi} = \delta_{\rm fi} -\frac{(2\pi)^4 i 
\delta (E_{\rm f} - E_{\rm i}) \delta^3 (\vec{P}_{\rm f} - \vec{P}_{\rm i})}
{V^2\sqrt{2E_A2E_B2E_C2E_D}}
\left( {\cal M}^{\rm post}_{q_1\bar{q}_1}
+{\cal M}^{\rm post}_{\bar{q}_2 q_2}
+{\cal M}^{\rm post}_{q_1q_2}+{\cal M}^{\rm post}_{\bar{q}_1\bar{q}_2} 
\right),
\end{equation}
for scattering in the post form, where $E_j$ is the energy of meson $j$;
$V$ is the volume in which every meson wave function is normalized;
$E_{\rm i}$ and $\vec{P}_{\rm i}$ ($E_{\rm f}$ and 
$\vec{P}_{\rm f}$) are the total energy and the total three-dimensional
momentum of the two initial (final) mesons, respectively;
${\cal M}^{\rm prior}_{q_1\bar{q}_2}$, 
${\cal M}^{\rm prior}_{\bar{q}_1 q_2}$, ${\cal M}^{\rm prior}_{q_1q_2}$, 
${\cal M}^{\rm prior}_{\bar{q}_1\bar{q}_2}$, 
${\cal M}^{\rm post}_{q_1\bar{q}_1}$, ${\cal M}^{\rm post}_{\bar{q}_2 q_2}$, 
${\cal M}^{\rm post}_{q_1q_2}$, and ${\cal M}^{\rm post}_{\bar{q}_1\bar{q}_2}$
are transition amplitudes. ${\cal M}_{q_1\bar {q}_2}^{\rm prior}$ and 
${\cal M}_{q_1\bar {q}_1}^{\rm post}$ are given by 
\begin{eqnarray}
{\cal M}_{q_1\bar {q}_2}^{\rm prior} & = &
\sqrt {2E_A2E_B2E_C2E_D}
\int \frac {d^3 p_{q_1\bar {q}_2}}{(2\pi)^3} 
     \frac {d^3 p_{q_2\bar {q}_1}}{(2\pi)^3}      \nonumber    \\
& & \psi^+_{q_1\bar {q}_2} (\vec {p}_{q_1\bar {q}_2}) 
\psi^+_{q_2\bar {q}_1} (\vec {p}_{q_2\bar {q}_1}) V_{q_1\bar {q}_2}
\psi_{q_1\bar {q}_1} (\vec {p}_{q_1\bar {q}_1}) 
\psi_{q_2\bar {q}_2} (\vec {p}_{q_2\bar {q}_2}),    
\end{eqnarray}
\begin{eqnarray}
{\cal M}_{q_1\bar {q}_1}^{\rm post} & = & 
\sqrt {2E_A2E_B2E_C2E_D}
\int \frac {d^3 p_{q_1\bar {q}_1}}{(2\pi)^3} 
     \frac {d^3 p_{q_2\bar {q}_2}}{(2\pi)^3}    \nonumber   \\
& & \psi^+_{q_1\bar {q}_2} (\vec {p}_{q_1\bar {q}_2}) 
\psi^+_{q_2\bar {q}_1} (\vec {p}_{q_2\bar {q}_1}) V_{q_1\bar {q}_1}
\psi_{q_1\bar {q}_1} (\vec {p}_{q_1\bar {q}_1}) 
\psi_{q_2\bar {q}_2} (\vec {p}_{q_2\bar {q}_2}),
\end{eqnarray}
where $\vec{p}_{ab}$ is the relative momentum of constituents $a$ and $b$;
$\psi_{ab}(\vec{p}_{ab})$ is the mesonic quark-antiquark wave function which
is the product of the color wave function, the flavor wave function, the spin 
wave function, and the relative-motion wave function of $a$ and $b$; 
$\psi_{ab}^+$ is the Hermitean conjugate of $\psi_{ab}$.
${\cal M}_{\bar {q}_1 q_2}^{\rm prior}$ (${\cal M}_{q_1 q_2}^{\rm prior}$, 
${\cal M}_{\bar {q}_1 \bar {q}_2}^{\rm prior}$) is obtained by replacing
$V_{q_1\bar {q}_2}$ in Eq. (12) with $V_{\bar {q}_1 q_2}$ ($V_{q_1 q_2}$, 
$V_{\bar {q}_1 \bar {q}_2}$), and ${\cal M}_{\bar {q}_2q_2}^{\rm post}$ 
(${\cal M}_{q_1 q_2}^{\rm post}$, 
${\cal M}_{\bar {q}_1 \bar {q}_2}^{\rm post}$) by replacing 
$V_{q_1\bar {q}_1}$ in Eq. (13) with $V_{\bar {q}_2q_2}$ ($V_{q_1 q_2}$, 
$V_{\bar {q}_1 \bar {q}_2}$).
$V_{ab}$ is the potential between $a$ and $b$ in momentum space, and is the 
Fourier transform of the following expression in coordinate space \cite{JSX}:
\begin{eqnarray}
V_{ab}(\vec {r}) = V_{\rm{si}}(\vec {r})+V_{\rm{ss}}(\vec {r}),
\end{eqnarray}
where $\vec {r}$ is the relative coordinate of $a$
and $b$, $V_{\rm{si}}$ the central spin-independent potential, and
$V_{\rm{ss}}$ the spin-spin interaction. 

The spin-independent potential depends on temperature $T$ and below the 
critical temperature $T_{\rm c}=0.175$ GeV is given by
\begin{equation}
V_{\rm {si}}(\vec {r})=
-\frac {\vec {\lambda}_a}{2} \cdot \frac {\vec{\lambda}_b}{2}
\xi_1 \left[ 1.3- \left( \frac {T}{T_{\rm c}} \right)^4 \right]
\tanh (\xi_2 r)
+ \frac {\vec {\lambda}_a}{2} \cdot \frac {\vec {\lambda}_b}{2}
\frac {6\pi}{25} \frac {v(\lambda r)}{r} \exp (-\xi_3 r),
\end{equation}
where $\xi_1 =0.525$ GeV, $\xi_2 =1.5[0.75+0.25 (T/{T_{\rm c}})^{10}]^6$
GeV, $\xi_3 =0.6$ GeV, and $\lambda=\sqrt{3b_0/16\pi^2 \alpha'}$ in which
$\alpha'=1.04$ GeV$^{-2}$ and $b_{0}=11-\frac{2}{3}N_{f}$ with the
quark flavor number $N_{f}=4$. $\vec {\lambda}_a$ are the
Gell-Mann matrices for the color generators of constituent quark
or antiquark labeled as $a$. The dimensionless function $v(x)$ is
given by Buchm\"{u}ller and Tye \cite{BT}. The short-distance part
of the spin-independent potential originates from one-gluon
exchange plus perturbative one- and two-loop corrections. The
intermediate-distance and large-distance part of the
spin-independent potential fits well the numerical potential which
was obtained in the lattice gauge calculations \cite{KLP}. At large
distances the spin-independent potential is independent of the
relative coordinate and obviously exhibits a plateau at $T/T_{\rm c} > 0.55$. 
The plateau height decreases with increasing
temperature. This means that confinement becomes weaker and
weaker.

Denote the spin and the mass of constituent $a$ by $\vec {s}_a$ and $m_a$,
respectively. The spin-spin
interaction with relativistic effects \cite{BS1992,GI,Xu2002} is
\begin{eqnarray}
V_{\rm ss}(\vec {r})=
- \frac {\vec {\lambda}_a}{2} \cdot \frac {\vec {\lambda}_b}{2}
\frac {16\pi^2}{25}\frac{d^3}{\pi^{3/2}}{\rm e}^{-d^2r^2} 
\frac {\vec {s}_a \cdot \vec
{s} _b} {m_am_b}
+ \frac {\vec {\lambda}_a}{2} \cdot \frac {\vec {\lambda}_b}{2}
  \frac {4\pi}{25} \frac {1} {r}
\frac {d^2v(\lambda r)}{dr^2} \frac {\vec {s}_a \cdot \vec {s}_b}{m_am_b},
\end{eqnarray}
with
\begin{eqnarray}
d^2=d_1^2\left[\frac{1}{2}+\frac{1}{2}
\left(\frac{4m_a m_b}{(m_a+m_b)^2}\right)^4\right]
+d_2^2\left(\frac{2m_am_b}{m_a+m_b}\right)^2,
\end{eqnarray}
where $d_1=0.15$ GeV and $d_2=0.705$.

$V_{ab}(\vec{r})$ at $T=0$ provides a quark potential in vacuum, and for
$T>0.6T_{\rm c}$ does so in hadronic matter that results from the quark-gluon
plasma. Meson masses and mesonic quark-antiquark relative-motion wave functions
in coordinate space are determined from the Schr\"odinger equation with
$V_{ab}(\vec {r})$. They depend on temperature.
The first term given in Eq. (15) stands for the confining potential. In the
confinement regime the mesonic quark-antiquark relative-motion wave functions
mainly determined by the confining potential are nonperturbative.
At low energies near threshold of $A(q_1\bar{q}_1)+B(q_2\bar{q}_2) \to 
C(q_2\bar{q}_1)+D(q_1\bar{q}_2)$ the nonperturbative part of the
$q_1\bar{q}_1$ and $q_2\bar{q}_2$ wave functions must overlap. 
Even though the distance of $q_2$ and $\bar{q}_1$ ($q_1$ and $\bar{q}_2$)
is large, the nonperturbative correlation corresponding to the nonperturbative
part leads to formation of a bound state of $q_2$ and $\bar{q}_1$
($q_1$ and $\bar{q}_2$) during the hadronization of $q_1$, $\bar{q}_1$,
$q_2$, and $\bar{q}_2$, i.e., meson $C$ ($D$).

Since the sum of the spin operators of $q_1$, $\bar{q}_1$, $q_2$, and 
$\bar{q}_2$ commutes with the potential $V_{ab}$, the total spin of 
the two final mesons equals the one of the two initial mesons. 
The following reactions have been considered in Ref. \cite{SX}:
\begin{eqnarray}
\nonumber
I=2~\pi\pi \to \rho\rho, \quad I=1~KK \to K^* K^*, \quad I=1~KK^* \to K^*K^*,
\\  \nonumber
I=3/2~\pi K \to \rho K^*, \quad I=3/2~\pi K^* \to \rho K^*, 
\\  \nonumber
I=3/2~\rho K \to \rho K^*, \quad I=3/2~\pi K^* \to \rho K.
\end{eqnarray}
The seven isospin channels of these reactions are endothermic.
The unpolarized cross section is the average of the one for scattering in
the prior form and the one for scattering in the post form. Denote by
$\sqrt s$ the center-of-mass energy of the two initial mesons. 
While $\sqrt s$ increases from threshold, 
the cross section for each isospin channel
rises very rapidly from 0, arrives at a maximum value
(peak cross section), and decreases rapidly. The cross sections for these 
reactions change considerably with the temperature of hadronic matter. For 
example,
every reaction has a rising peak cross section when the temperature becomes
critical.

\section{Meson-Meson Reactions due to Quark-Antiquark Annihilation and 
Creation}

A quark in an initial meson and an antiquark in another initial meson
annihilate into a gluon, and subsequently the gluon creates another 
quark-antiquark pair. This quark-antiquark annihilation and creation causes
two-to-two meson-meson scattering, $A(q_1\bar{q}_1)+B(q_2\bar{q}_2) \to
C(q_3\bar{q}_1)+D(q_2\bar{q}_4)$ from $q_1+\bar{q}_2 \to q_3+\bar{q}_4$ and
$A(q_1\bar{q}_1)+B(q_2\bar{q}_2) \to C(q_1\bar{q}_4)+D(q_3\bar{q}_2)$ from 
$\bar{q}_1+q_2 \to q_3+\bar{q}_4$. The $S$-matrix element for $A+B \to C+D$
is \cite{SXW}
\begin{equation}
S_{\rm fi} =  \delta_{\rm fi} -(2\pi)^4 i \delta (E_{\rm f} - E_{\rm i})
\delta^3 (\vec{P}_{\rm f} - \vec{P}_{\rm i})
\frac {{\cal M}_{{\rm a}q_1\bar{q}_2}+{\cal M}_{{\rm a}\bar{q}_1 q_2}}
{V^2\sqrt{2E_A2E_B2E_C2E_D}},
\end{equation}
where ${\cal M}_{{\rm a}q_1\bar{q}_2}$ and ${\cal M}_{{\rm a}\bar{q}_1 q_2}$ 
are the transition amplitudes that may be written as:
\begin{eqnarray}
{\cal M}_{{\rm a}q_1\bar {q}_2} & = &
\sqrt {2E_A2E_B2E_C2E_D}
\int \frac {d^3p_{q_1\bar{q}_1}}{(2\pi)^3}\frac {d^3p_{q_2\bar{q}_2}}{(2\pi)^3}
                       \nonumber         \\
& &
\psi^+_{q_3\bar {q}_1} (\vec {p}_{q_3\bar {q}_1})
\psi^+_{q_2\bar {q}_4} (\vec {p}_{q_2\bar {q}_4})
V_{{\rm a}q_1\bar{q}_2}
\psi_{q_1\bar {q}_1} (\vec {p}_{q_1\bar {q}_1})
\psi_{q_2\bar {q}_2} (\vec {p}_{q_2\bar {q}_2}),
\end{eqnarray}
\begin{eqnarray}
{\cal M}_{{\rm a}\bar{q}_1q_2} & = &
\sqrt {2E_A2E_B2E_C2E_D}
\int \frac {d^3p_{q_1\bar{q}_1}}{(2\pi)^3}\frac {d^3p_{q_2\bar{q}_2}}{(2\pi)^3}
                       \nonumber         \\
& &
\psi^+_{q_1\bar {q}_4} (\vec {p}_{q_1\bar {q}_4})
\psi^+_{q_3\bar {q}_2} (\vec {p}_{q_3\bar {q}_2})
V_{{\rm a}\bar{q}_1q_2}
\psi_{q_1\bar {q}_1} (\vec {p}_{q_1\bar {q}_1})
\psi_{q_2\bar {q}_2} (\vec {p}_{q_2\bar {q}_2}),
\end{eqnarray}
where $V_{{\rm a}q_1\bar{q}_2}$ and $V_{{\rm a}\bar{q}_1q_2}$ are the 
transition potentials for $q_1+\bar{q}_2 \to q_3+\bar{q}_4$ and
$\bar{q}_1+q_2 \to q_3+\bar{q}_4$, respectively. The transition potentials
have been derived in perturbative quantum chromodynamics (QCD).
Since the commutators of the transition potentials and the
sum of the four constituent spin operators do
not equal zero, the total spin of the two final mesons may not equal the
one of the two initial mesons \cite{WX}. The following reactions have
been considered in Refs. \cite{SXW,WX}:
\begin{eqnarray}
\nonumber
\pi \pi \to K \bar K, \quad \pi \pi \to K \bar{K}^\ast, 
\quad \pi \pi \to K^\ast \bar{K}, \quad \pi \pi \to K^\ast \bar{K}^\ast,
\\  \nonumber
\pi \rho \to K \bar{K}, \quad \pi \rho \to K \bar {K}^\ast, 
\quad \pi \rho \to K^* \bar{K}, \quad \pi \rho \to K^\ast \bar{K}^\ast,
\\  \nonumber
\rho \rho \to K^\ast \bar{K}^\ast, \quad K \bar {K} \to \rho \rho,
\quad K \bar{K}^\ast \to \rho \rho, \quad K^* \bar{K} \to \rho \rho,
\\  \nonumber
K \bar {K} \to K \bar {K}^\ast , \quad K \bar{K} \to K^* \bar{K},
\quad K \bar {K} \to K^* \bar {K}^\ast , 
\\  \nonumber
K \bar{K}^\ast \to K^* \bar{K}^\ast, \quad K^\ast \bar{K} \to K^* \bar{K}^\ast,
\\  \nonumber
\pi K \to \pi K^\ast, \quad \pi K \to \rho K,
\\  \nonumber
I=1~ \pi \pi \to \rho \rho.
\end{eqnarray}

In some regimes these reactions receive contributions from quark-antiquark 
annihilation
and creation. Unpolarized cross sections for these reactions depend on $T$ 
and $\sqrt s$. The temperature dependence arises
from the meson masses and the quark-antiquark relative-motion wave functions
of the initial and final mesons. With increasing $\sqrt s$
from the threshold energy, the cross section for any endothermic reaction of 
these reactions at a given temperature increases very rapidly to a peak
cross section first, and then decreases or displays a plateau for some
energy interval.

\section{Meson-Meson Reactions due to Quark-Interchange as well as 
Quark-Antiquark Annihilation and Creation}	

Transition amplitudes are proportional to flavor matrix elements. Flavor
matrix elements are calculated with flavor wave functions of initial and final
mesons. If the flavor matrix element of a reaction is zero, the reaction can
not occur. The flavor matrix element depends on the total isospin of the two
initial or final mesons and the mechanism (quark interchange, quark-antiquark 
annihilation and creation) which causes the reaction. 
From the flavor matrix elements of
$\pi\pi \to \rho\rho$ we know that $\pi\pi \to \rho\rho$ for $I=2$ is governed
by quark interchange, $\pi\pi \to \rho\rho$ for $I=1$ by
quark-antiquark annihilation and creation, and $\pi\pi \to \rho\rho$ for $I=0$
by quark interchange as well as quark-antiquark annihilation and
creation. Even though the four reactions
$\pi K \to \rho K^*$, $\pi K^* \to \rho K$, $\pi K^* \to \rho K^*$, and
$\rho K \to \rho K^*$ are governed by quark interchange when $I=3/2$,
they are governed by quark interchange as well as quark-antiquark annihilation
and creation when $I=1/2$. Therefore, some isospin channels of reactions are
governed by quark interchange as well as quark-antiquark annihilation
and creation. The $S$-matrix element for this kind of scattering is \cite{SXW}
\begin{eqnarray}
S_{\rm fi} & = & \delta_{\rm fi} -\frac{(2\pi)^4 i 
\delta (E_{\rm f} - E_{\rm i}) \delta^3 (\vec{P}_{\rm f} - \vec{P}_{\rm i})}
{V^2\sqrt{2E_A2E_B2E_C2E_D}}       \nonumber \\
& & \left( {\cal M}^{\rm prior}_{q_1\bar{q}_2}
+{\cal M}^{\rm prior}_{\bar{q}_1 q_2}
+{\cal M}^{\rm prior}_{q_1q_2}+{\cal M}^{\rm prior}_{\bar{q}_1\bar{q}_2}
+{\cal M}_{{\rm a}q_1\bar{q}_2}+{\cal M}_{{\rm a}\bar{q}_1 q_2}
\right) ,
\end{eqnarray}
with the quark-interchange process in the prior form or
\begin{eqnarray}
S_{\rm fi} & = & \delta_{\rm fi} -\frac{(2\pi)^4 i 
\delta (E_{\rm f} - E_{\rm i}) \delta^3 (\vec{P}_{\rm f} - \vec{P}_{\rm i})}
{V^2\sqrt{2E_A2E_B2E_C2E_D}}       \nonumber \\
& & \left( {\cal M}^{\rm post}_{q_1\bar{q}_1}
+{\cal M}^{\rm post}_{\bar{q}_2 q_2}
+{\cal M}^{\rm post}_{q_1q_2}+{\cal M}^{\rm post}_{\bar{q}_1\bar{q}_2}
+{\cal M}_{{\rm a}q_1\bar{q}_2}+{\cal M}_{{\rm a}\bar{q}_1 q_2}
\right),
\end{eqnarray}
with the quark-interchange process in the post form. 

Unpolarized cross sections for $\pi\pi \to \rho\rho$ for $I=0$ have been
obtained in Ref. \cite{SXW} and ones for $\pi K \to \rho K^*$ for $I=1/2$,
$\pi K^* \to \rho K$ for $I=1/2$, $\pi K^* \to \rho K^\ast$ for $I=1/2$,
and $\rho K \to \rho K^*$ for $I=1/2$ have been 
obtained in Ref. \cite{YXW}.
Near threshold quark interchange dominates the five channels near
the critical temperature; in the other energy region quark-antiquark
annihilation and creation may dominate the five channels.

\section{Meson-Meson Reactions from Quark-Antiquark Annihilation}	

A quark in one initial meson and an antiquark in the other initial meson 
annihilate into a gluon, and subsequently the gluon is absorbed by the other
antiquark or quark. This quark-antiquark annihilation leads to two-to-one 
meson-meson scattering. The reaction $A(q_1\bar{q}_1)+B(q_2\bar{q}_2) \to 
H(q_2\bar{q}_1)$ is caused by $q_1+\bar{q}_2+\bar{q}_1 \to \bar{q}_1$ and
$q_1+\bar{q}_2+q_2 \to q_2$, and $A(q_1\bar{q}_1)+B(q_2\bar{q}_2) \to 
H(q_1\bar{q}_2)$ is caused by $q_2+\bar{q}_1+q_1 \to q_1$ and
$q_2+\bar{q}_1+\bar{q}_2 \to \bar{q}_2$. The $S$-matrix element for $A+B \to H$
is \cite{YXW}
\begin{equation}
S_{\rm fi} = \delta_{\rm fi} -(2\pi)^4 i \delta (E_{\rm f} - E_{\rm i})
\delta^3 (\vec{P}_{\rm f} - \vec{P}_{\rm i})
\frac {{\cal M}_{{\rm r}q_1\bar{q}_2 \bar{q}_1}+
{\cal M}_{{\rm r}q_1\bar{q}_2 q_2}+{\cal M}_{{\rm r}q_2 \bar{q}_1 q_1}
+{\cal M}_{{\rm r}q_2 \bar{q}_1 \bar{q}_2}}
{V^{\frac {3}{2}}\sqrt{2E_A2E_B2E_H}},
\end{equation}
where ${\cal M}_{{\rm r}q_1\bar{q}_2 \bar{q}_1}$, 
${\cal M}_{{\rm r}q_1\bar{q}_2 q_2}$, ${\cal M}_{{\rm r}q_2 \bar{q}_1 q_1}$,
and ${\cal M}_{{\rm r}q_2 \bar{q}_1 \bar{q}_2}$ are 
the transition amplitudes that may be written as
\begin{eqnarray}
{\cal M}_{{\rm r}q_1\bar{q}_2 \bar{q}_1}
& = &
\sqrt {2E_A2E_B2E_H}
\int \frac{d^3 p_{q_1\bar{q}_1}}{(2\pi)^3}\frac{d^3 p_{q_2\bar{q}_2}}{(2\pi)^3}
\psi_{q_2\bar{q}_1}^+ (\vec {p}_{q_2\bar{q}_1})
V_{{\rm r}q_1\bar{q}_2 \bar{q}_1}
    \nonumber   \\
& &
\psi_{q_1\bar{q}_1} (\vec {p}_{q_1\bar {q}_1})
\psi_{q_2\bar{q}_2} (\vec {p}_{q_2\bar {q}_2}),
\end{eqnarray}
\begin{eqnarray}
{\cal M}_{{\rm r}q_1\bar{q}_2 q_2}
& = &
\sqrt {2E_A2E_B2E_H}
\int \frac{d^3 p_{q_1\bar{q}_1}}{(2\pi)^3}
\frac{d^3 p_{q_2\bar{q}_2}}{(2\pi)^3}
\psi_{q_2\bar{q}_1}^+ (\vec {p}_{q_2\bar{q}_1})V_{{\rm r}q_1\bar{q}_2 q_2}
    \nonumber   \\
& &
\psi_{q_1\bar{q}_1} (\vec {p}_{q_1\bar {q}_1})
\psi_{q_2\bar{q}_2} (\vec {p}_{q_2\bar {q}_2}),
\end{eqnarray}
\begin{eqnarray}
{\cal M}_{{\rm r}q_2 \bar{q}_1 q_1}
& = &
\sqrt {2E_A2E_B2E_H}
\int \frac{d^3 p_{q_1\bar{q}_1}}{(2\pi)^3}\frac{d^3 p_{q_2\bar{q}_2}}{(2\pi)^3}
\psi_{q_1\bar{q}_2}^+ (\vec {p}_{q_1\bar{q}_2})V_{{\rm r}q_2 \bar{q}_1 q_1}
    \nonumber   \\
& &
\psi_{q_1\bar{q}_1} (\vec {p}_{q_1\bar {q}_1})
\psi_{q_2\bar{q}_2} (\vec {p}_{q_2\bar {q}_2}),
\end{eqnarray}
\begin{eqnarray}
{\cal M}_{{\rm r}q_2 \bar{q}_1 \bar{q}_2}
& = &
\sqrt {2E_A2E_B2E_H}
\int \frac{d^3 p_{q_1\bar{q}_1}}{(2\pi)^3}\frac{d^3 p_{q_2\bar{q}_2}}{(2\pi)^3}
\psi_{q_1\bar{q}_2}^+ (\vec {p}_{q_1\bar{q}_2})
V_{{\rm r} q_2\bar{q}_1 \bar{q}_2}
    \nonumber   \\
& &
\psi_{q_1\bar{q}_1} (\vec {p}_{q_1\bar {q}_1})
\psi_{q_2\bar{q}_2} (\vec {p}_{q_2\bar {q}_2}),
\end{eqnarray}
where $V_{{\rm r}q_1\bar{q}_2\bar{q}_1}$ 
($V_{{\rm r}q_1\bar{q}_2q_2}$) represents the transition potential for the
annihilation of $q_1$ and $\bar{q}_2$ into a gluon and the subsequent
absorption of the gluon by $\bar{q}_1$ in meson $A$ ($q_2$ in meson $B$);
$V_{{\rm r}q_2\bar{q}_1q_1}$ ($V_{{\rm r}q_2\bar{q}_1\bar{q}_2}$) represents
the transition potential for the annihilation of $q_2$ and $\bar{q}_1$ into 
a gluon and the subsequent absorption of the gluon by $q_1$ in meson $A$ 
($\bar{q}_2$ in meson $B$). The transition potentials have been derived from
the Feynman rules in perturbative QCD.

\vspace{0.2in}
\begin{figure}[ph]
\centerline{\includegraphics[width=3.5in]{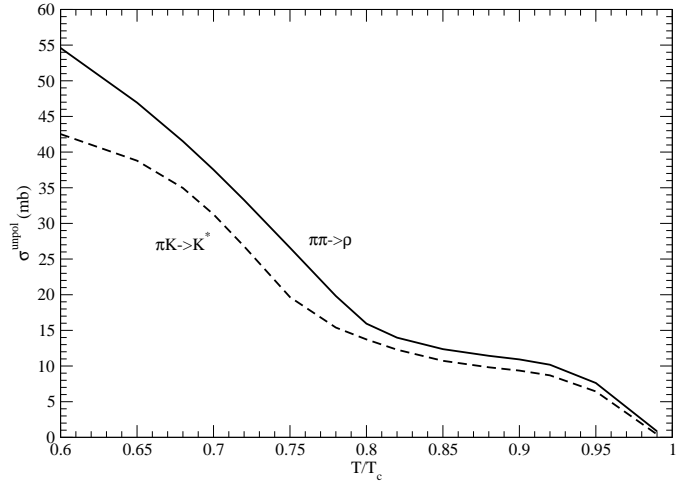}}
\vspace*{8pt}
\caption{Unpolarized cross sections for $\pi\pi \to \rho$ and $\pi K \to 
K^\ast$ as functions of $T/T_{\rm c}$.\protect\label{fig1}}
\end{figure}

The transition amplitudes result in cross sections for $\pi\pi \to \rho$ and 
$\pi K \to K^\ast$ in vacuum, which agree with empirical data \cite{YXW}. 
Unpolarized
cross sections for $0.6 \leq T/T_{\rm c} <1$ are shown in Fig. 1. The weakening
confinement with increasing temperature makes combining the final quark and
the final antiquark into a meson more difficult, and thus reduces the cross
sections. The temperature dependence of the cross sections is remarkable.

\section{Meson-Meson Reactions from Quark-Antiquark Creation}	

In the collision of mesons $A(q_1\bar{q}_1)$ and $B(q_2\bar{q}_2)$ a 
constituent quark or antiquark may emit a virtual gluon which subsequently
splits into quark $q_3$ and antiquark $\bar{q}_4$. If $\bar{q}_4$ and $q_1$
combine into a meson, then $q_3$ and $\bar{q}_2$ combine into another meson. If
$\bar{q}_4$ and $q_2$ combine into a meson, then $q_3$ and $\bar{q}_1$ combine 
into another meson. The quark-antiquark creation thus leads to the reaction
$A(q_1\bar{q}_1)+B(q_2\bar{q}_2) \to C_1(q_1\bar{q}_4)+C_2(q_2\bar{q}_1)
+C_3(q_3\bar{q}_2)$ and $A(q_1\bar{q}_1)+B(q_2\bar{q}_2) \to 
C_1(q_1\bar{q}_2)+C_2(q_2\bar{q}_4)+C_3(q_3\bar{q}_1)$. 
The $S$-matrix element for $A+B \to C_1+C_2+C_3$ is \cite{LXW}
\begin{eqnarray}
S_{\rm fi} & = & \delta_{\rm fi} - \frac{ (2\pi)^4 i \delta 
(E_{\rm f} - E_{\rm i})\delta^3 (\vec{P}_{\rm f} - \vec{P}_{\rm i}) }
{\sqrt {V^5}\sqrt{2E_A2E_B2E_{C_1}2E_{C_2}2E_{C_3}}}        \nonumber \\
& & \left( {\cal M}_{{\rm D}_1}+{\cal M}_{{\rm D}_2}+{\cal M}_{{\rm D}_3}
+{\cal M}_{{\rm D}_4}+{\cal M}_{{\rm D}_5}+{\cal M}_{{\rm D}_6}
+{\cal M}_{{\rm D}_7}+{\cal M}_{{\rm D}_8} \right),
\end{eqnarray}
where ${\cal M}_{{\rm D}_h}$ ($h=1, \cdot \cdot \cdot, 8$) are transition 
amplitudes. ${\cal M}_{{\rm D}_1}$ and ${\cal M}_{{\rm D}_5}$ may be written
as 
\begin{eqnarray}
{\cal M}_{{\rm D}_1}
& = &
\sqrt {2E_A2E_B2E_{C_1}2E_{C_2}2E_{C_3}}
\int \frac{d^3 p_{q_2\bar{q}_2}}{(2\pi)^3}\frac{d^3 p_{q_1\bar{q}_4}}{(2\pi)^3}
      \nonumber   \\
& &
\psi_{q_1\bar{q}_4}^+ (\vec {p}_{q_1\bar{q}_4})
\psi_{q_2\bar{q}_1}^+ (\vec {p}_{q_2\bar{q}_1})
\psi_{q_3\bar{q}_2}^+ (\vec {p}_{q_3\bar{q}_2})V_{{\rm D}_1}
\psi_{q_1\bar{q}_1} (\vec {p}_{q_1\bar {q}_1})
\psi_{q_2\bar{q}_2} (\vec {p}_{q_2\bar {q}_2}),
\end{eqnarray}
\begin{eqnarray}
{\cal M}_{{\rm D}_5}
& = &
\sqrt {2E_A2E_B2E_{C_1}2E_{C_2}2E_{C_3}}
\int \frac{d^3 p_{q_1\bar{q}_1}}{(2\pi)^3}\frac{d^3 p_{q_2\bar{q}_2}}{(2\pi)^3}
      \nonumber   \\
& &
\psi_{q_1\bar{q}_2}^+ (\vec {p}_{q_1\bar{q}_2})
\psi_{q_2\bar{q}_4}^+ (\vec {p}_{q_2\bar{q}_4})
\psi_{q_3\bar{q}_1}^+ (\vec {p}_{q_3\bar{q}_1})V_{{\rm D}_5}
\psi_{q_1\bar{q}_1} (\vec {p}_{q_1\bar {q}_1})
\psi_{q_2\bar{q}_2} (\vec {p}_{q_2\bar {q}_2}).
\end{eqnarray}
${\cal M}_{{\rm D}_2}$ (${\cal M}_{{\rm D}_3}$, ${\cal M}_{{\rm D}_4}$) is
obtained by replacing $d^3 p_{q_2\bar{q}_2}d^3 p_{q_1\bar{q}_4}V_{{\rm D}_1}$ 
in Eq. (29) with $d^3 p_{q_1\bar{q}_1}d^3 p_{q_2\bar{q}_2}V_{{\rm D}_2}$ 
($d^3 p_{q_1\bar{q}_1}d^3 p_{q_2\bar{q}_2}V_{{\rm D}_3}$, 
$d^3 p_{q_1\bar{q}_1}d^3 p_{q_3\bar{q}_2}V_{{\rm D}_4}$), and
${\cal M}_{{\rm D}_6}$ (${\cal M}_{{\rm D}_7}$, ${\cal M}_{{\rm D}_8}$) by
replacing $d^3 p_{q_1\bar{q}_1}d^3 p_{q_2\bar{q}_2}V_{{\rm D}_5}$ in Eq. (30)
with $d^3 p_{q_2\bar{q}_2}d^3 p_{q_3\bar{q}_1}V_{{\rm D}_6}$ 
($d^3 p_{q_1\bar{q}_1}d^3 p_{q_2\bar{q}_4}V_{{\rm D}_7}$, 
$d^3 p_{q_1\bar{q}_1}d^3 p_{q_2\bar{q}_2}V_{{\rm D}_8}$).
$V_{{\rm D}_1}$ ($V_{{\rm D}_2}$, $V_{{\rm D}_3}$, $V_{{\rm D}_4}$) 
represents the transition potential for $q_1 \to q_1+q_3+\bar{q}_4$
($\bar{q}_1 \to \bar{q}_1+q_3+\bar{q}_4$, $q_2 \to q_2+q_3+\bar{q}_4$,
$\bar{q}_2 \to \bar{q}_2+q_3+\bar{q}_4$) in 
$A(q_1\bar{q}_1)+B(q_2\bar{q}_2) \to C_1(q_1\bar{q}_4)+C_2(q_2\bar{q}_1)
+C_3(q_3\bar{q}_2)$, and 
$V_{{\rm D}_5}$ ($V_{{\rm D}_6}$, $V_{{\rm D}_7}$, $V_{{\rm D}_8}$) 
indicates the transition potential for $q_1 \to q_1+q_3+\bar{q}_4$
($\bar{q}_1 \to \bar{q}_1+q_3+\bar{q}_4$, $q_2 \to q_2+q_3+\bar{q}_4$,
$\bar{q}_2 \to \bar{q}_2+q_3+\bar{q}_4$) in
$A(q_1\bar{q}_1)+B(q_2\bar{q}_2) \to 
C_1(q_1\bar{q}_2)+C_2(q_2\bar{q}_4)+C_3(q_3\bar{q}_1)$.
The transition potentials have been derived in perturbative QCD.

The following two-to-three meson-meson reactions are considered in Ref. 
\cite{LXW}:
\begin{displaymath}
\pi \pi \to \pi K\bar{K},~\pi K \to \pi \pi K,~\pi K \to KK\bar{K},
~KK \to \pi KK,~K\bar{K} \to \pi K\bar{K}.
\end{displaymath}
Unpolarized cross sections for these reactions have been obtained from the
eight transition amplitudes.
At zero temperature the cross sections for $\pi K \to \pi\pi K$ for $I=3/2$ 
are near the experimental data. If the cross section for any reaction
at a given temperature has a maximum, then the peak cross section
decreases as the temperature approaches the critical
temperature. By comparison with inelastic
two-to-two meson-meson scattering, we find that two-to-three meson-meson 
scattering may be as important as inelastic two-to-two meson-meson scattering.

\section{Multi Mesons produced in Meson-Meson Reactions}	

Meson-meson scattering is usually governed by a single Feynman diagram each, 
i.e. one transition amplitude corresponds to one Feynman diagram at tree level.
Denote inelastic
meson-meson scattering by $A(q_1\bar{q}_1)
+B(q_2\bar{q}_2) \to C_1(q_1^{\prime}\bar{q}_1^{\prime})+ \cdot \cdot
\cdot +C_{\rm n}(q_{\rm n}^{\prime}\bar{q}_{\rm n}^{\prime})$, and let 
${\rm D}_1, \cdot \cdot \cdot , {\rm D}_{\rm m}$
stand for the Feynman diagrams.
For diagram ${\rm D}_i$ the transition amplitude is written as
\begin{eqnarray}
{\cal M}_{{\rm D}_i}
& = &
\sqrt {2E_A2E_B2E_{C_1} \cdot \cdot \cdot 2E_{C_{\rm n}}}
\int \frac{d^3 p_{a\bar b}}{(2\pi)^3}\frac{d^3 p_{c\bar d}}{(2\pi)^3}
      \nonumber   \\
& &
\psi_{q_1^{\prime}\bar{q}_1^{\prime}}^+ 
(\vec{p}_{q_1^{\prime}\bar{q}_1^{\prime}})
\cdot \cdot \cdot
\psi_{q_{\rm n}^{\prime}\bar{q}_{\rm n}^{\prime}}^+
(\vec {p}_{q_{\rm n}^{\prime}\bar{q}_{\rm n}^{\prime}}) V_{{\rm D}_i}
\psi_{q_1\bar{q}_1} (\vec {p}_{q_1\bar {q}_1})
\psi_{q_2\bar{q}_2} (\vec {p}_{q_2\bar {q}_2}),
\end{eqnarray}
where $p_{a\bar b}$ and $p_{c\bar d}$ are two of the mesonic quark-antiquark
relative momenta, and $V_{{\rm D}_i}$ is the transition potential corresponding
to diagram ${\rm D}_i$.
The expression holds true for two-to-one, two-to-two, and
two-to-three meson-meson scattering, but is assumed to be true for
two-to-four meson-meson scattering, two-to-five meson-meson scattering, 
and so on. Let $m_A$ ($m_B$) be the mass of meson $A$ ($B$), and let $P_j$
and $J_j$ ($j=A,B,C_1, \cdot \cdot \cdot, C_{\rm n}$) be the four-momentum
and the angular momentum of meson $j$ with
the magnetic projection quantum number $J_{jz}$, respectively. The
unpolarized cross section for $A+B \to C_1+ \cdot \cdot \cdot +C_{\rm n}$ is
\begin{eqnarray}
\sigma^{\rm unpol} (\sqrt{s},T) & = &
\frac {1}{(2J_A+1)(2J_B+1)}
\frac {(2\pi)^4}{4\sqrt {(P_A \cdot P_B)^2 - m_A^2m_B^2}}
      \nonumber  \\
& & \int \frac {d^3P_{C_1}}{(2\pi)^3 2E_{C_1}} \cdot \cdot \cdot
\frac {d^3P_{C_{\rm n}}}{(2\pi)^3 2E_{C_{\rm n}}}
\delta (E_{\rm f}-E_{\rm i}) \delta^3 (\vec{P}_{\rm f}-\vec{P}_{\rm i})
      \nonumber  \\
& & \sum\limits_{J_{Az}J_{Bz}J_{C_1z} \cdot \cdot \cdot J_{{C_{\rm n}z}}}
\mid {\cal M}_{{\rm D}_1}+ \cdot \cdot \cdot +{{\cal M}_{{\rm D}_{\rm m}}}
\mid^2 ,
\end{eqnarray}

\section{Summary}	

We have seen that in solving physical problems precise parametrizations
of scattering amplitudes of quite a few reactions have been given, and the
$T$-matrix elements for inelastic meson-meson scattering are obtained from
the Lippmann-Schwinger equations or by the generalization
of the inverse amplitude method to multi channels. Starting from the 
$\chi$PT Lagrangians, the coupled-channel unitary approaches and the inverse
amplitude method offer more insights into the strong interaction through
meson-meson scattering. However, the approaches and the method based on 
meson degrees of freedom remain in vacuum.

For two mesons in the ground-state pseudoscalar octet and the ground-state
vector nonet, their scattering may be caused at tree level by one or
two of the processes: quark interchange, quark-antiquark annihilation,
quark-antiquark creation, and quark-antiquark annihilation and creation.
The transition potentials corresponding to quark-antiquark annihilation,
quark-antiquark creation, and quark-antiquark annihilation and creation have
been derived in QCD. From these processes two-to-one meson-meson scattering,
two-to-two meson-meson scattering, and two-to-three meson-meson scattering
have been studied. The transition amplitudes are composed of the
transition potentials and the mesonic quark-antiquark wave functions of which
the relative-motion part is the solution of the
Schr\"odinger equation with the temperature-dependent quark potential. 
Unpolarized cross sections obtained from the transition amplitudes
depend not only on the center-of-mass energy of the two initial
mesons but also on the temperature of hadronic matter. From vacuum to hadronic
matter the cross sections change considerably, and the influence of hadronic
matter on inelastic meson-meson scattering is remarkable.

Finally, particular attention is paid to two current subjects revealed by
experiments at the LHC. One subject is the large
CP asymmetry in the charmless three-body decay modes of $B$ mesons. One needs
to answer how large
is the contribution of inelastic meson-meson scattering to the CP asymmetry,
and whether future experiments can give more new decay
modes in which the CP violation is influenced by inelastic meson-meson
scattering. Another subject is the strong interaction between hadronic matter
and a meson with a large momentum. For Pb-Pb collisions 
at $\sqrt {s_{NN}}=5.02$ TeV the CMS data \cite{CMS2018}
and the ATLAS data \cite{ATLAS2018} show that the prompt-$J/\psi$ nuclear
modification factor stays unchanged when the $J/\psi$ transverse momentum goes
from 8 GeV/$c$ to 18 GeV/$c$. The unchanged nuclear
modification factor is nontrivial! The mechanism where three or more mesons are
produced in a collision of a light meson and a charmonium plays a 
key role in understanding the unchanged nuclear modification factor \cite{JXW}.
Inelastic meson-meson scattering is certainly important to a high-momentum 
meson penetrating hadronic matter.

\section*{Acknowledgments}

This work was supported by the National Natural Science Foundation of China
under Grant No. 11175111.


\begin{thebibliography}{0}
\bibitem{OO1997} J. A. Oller and E. Oset, {\it Nucl. Phys. A} {\bf 620}, 438 
(1997).
\bibitem{DGL} I. V. Danilkin, L. I. R. Gil, and M. F. M. Lutz, {\it Phys. Lett.
B} {\bf 703}, 504 (2011).
\bibitem{GLMOR} Z.-H. Guo, L. Liu, U.-G. Mei{\ss}ner, J. A. Oller, and A. 
Rusetsky, {\it Phys. Rev. D} {\bf 95}, 054004 (2017).
\bibitem{OO1998} J. A. Oller and E. Oset, {\it Nucl. Phys. A} {\bf 629}, 739 
(1998).
\bibitem{DV} I. Danilkin and M. Vanderhaeghen, {\it Phys. Lett. B} {\bf 789},
366 (2019).
\bibitem{DDV} I. Danilkin, O. Deineka, and M. Vanderhaeghen, {\it Phys. Rev. D}
{\bf101}, 054008 (2020).
\bibitem{MO} U.-G. Mei{\ss}ner and J. A. Oller, {\it Nucl. Phys. A} {\bf 679},
671 (2001).
\bibitem{GPSCZ} F.-K. Guo, R.-G. Ping, P.-N. Shen, H.-C. Chiang, and B. S. Zou,
{\it Nucl. Phys. A} {\bf 773}, 78 (2006).
\bibitem{SLTO} S. Sakai, W.-H. Liang, G. Toledo, and E. Oset, {\it Phys. Rev. 
D} {\bf 101}, 014005 (2020).
\bibitem{LHCb1} LHCb Collaboration, {\it Phys. Rev. D} {\bf 90}, 112004 (2014).
\bibitem{LHCb2} LHCb Collaboration, {\it Phys. Rev. Lett.} {\bf 123}, 231802 
(2019).
\bibitem{LHCb3} LHCb Collaboration, {\it Phys. Rev. Lett.} {\bf 124}, 031801 
(2020).
\bibitem{LHCb4} LHCb Collaboration, {\it Phys. Rev. D} {\bf 101}, 012006 
(2020).
\bibitem{NBCFL} J. H. A. Nogueira, I. Bediaga, A. B. R. Cavalcante, 
T. Frederico, and O. Luorenco, {\it Phys. Rev. D} {\bf 92}, 054010 (2015).
\bibitem{KMR} P. Koch, B. M\"uller, and J. Rafelski, {\it Phys. Rep.} 
{\bf 142}, 167 (1986).
\bibitem{LXW} W.-X. Li, X.-M. Xu, and H. J. Weber, {\it Phys. Rev. D} 
{\bf 101}, 014025 (2020).
\bibitem{ATLAS1000} ATLAS Collaboration, {\it Phys. Lett. B} {\bf 790}, 108 
(2019).
\bibitem{CMS1000} CMS Collaboration, {\it JHEP} {\bf 10}, 138 (2018).
\bibitem{ALICE1000} ALICE Collaboration, {\it Phys. Lett. B} {\bf 788}, 166 
(2019).
\bibitem{GL1} J. Gasser and H. Leutwyler, {\it Ann. Phys.} {\bf 158}, 142 
(1984).
\bibitem{GL2} J. Gasser and H. Leutwyler, {\it Nucl. Phys. B} {\bf 250}, 465 
(1985).
\bibitem{HLPT} P. Herrera-Sikl\'ody, J. I. Latorre, P. Pascual, J. Taron,
{\it Nucl. Phys. B} {\bf 497}, 345 (1997).
\bibitem{KL} R. Kaiser and H. Leutwyler, {\it Eur. Phys. J. C} {\bf 17}, 623
(2000).
\bibitem{AOR} M. Albaladejo, J. A. Oller, and L. Roca, {\it Phys. Rev. D} 
{\bf 82}, 094019 (2010).
\bibitem{TDKJO} A. M. Torres, L. R. Dai, C. Koren, D. Jido, and E. Oset, 
{\it Phys. Rev. D} {\bf 85}, 014027 (2012).
\bibitem{WI} J. Weinstein and N. Isgur, {\it Phys. Rev. D} {\bf 41}, 2236 
(1990).
\bibitem{AO} M. Albaladejo and J. A. Oller, {\it Phys. Rev. Lett.} {\bf 101}, 
252002 (2008).
\bibitem{OOP1} J. A. Oller, E. Oset, and J. R. Pel\'aez, {\it Phys. Rev. Lett.}
{\bf 80}, 3452 (1998).
\bibitem{OOP2} J. A. Oller, E. Oset, and J. R. Pel\'aez, {\it Phys. Rev. D} 
{\bf 59}, 074001 (1999).
\bibitem{Barnes2002} T. Barnes, arXiv:hep-ph/0202157.
\bibitem{Barnes2003} T. Barnes, arXiv:hep-ph/0311102.
\bibitem{PR} J. R. Pel\'aez and A. Rodas, {\it Eur. Phys. J. C} {\bf 78}, 897 
(2018).
\bibitem{SBGKL} Y. S. Surovtsev, P. Byd\v{z}ovsk\'y, T. Gutsche, 
R. Kami\'nski, V. E. Lyubovitskij, and M. Nagy, {\it Phys. Rev. D} {\bf 97}, 
014009 (2018).
\bibitem{DEW} J. J. Dudek, R. G. Edwards, and D. J. Wilson, {\it Phys. Rev. D}
{\bf 93}, 094506 (2016).
\bibitem{LDHS} D. Lohse, J. W. Durso, K. Holinde, and J. Speth, {\it Nucl. 
Phys. A} {\bf 516}, 513 (1990).
\bibitem{BKWX} G. E. Brown, C. M. Ko, Z. G. Wu, and L. H. Xia, {\it Phys. Rev. 
C} {\bf 43}, 1881 (1991).
\bibitem{TKANN} A. M. Torres, K. P. Khemchandani, L. M. Abreu, F. S. Navarra,
and M. Nielsen, {\it Phys. Rev. D} {\bf 97}, 056001 (2018).
\bibitem{BS1992} T. Barnes and E. S. Swanson, {\it Phys. Rev. D} {\bf 46}, 131
(1992).
\bibitem{LX} Y.-Q. Li and X.-M. Xu, {\it Nucl. Phys. A} {\bf 794}, 210 (2007).
\bibitem{JSX} S.-T. Ji, Z.-Y. Shen, and X.-M. Xu, {\it J. Phys. G} {\bf 42}, 
095110 (2015).
\bibitem{BT} W. Buchm\"{u}ller and S.-H. H. Tye, {\it Phys. Rev. D} {\bf 24}, 
132 (1981).
\bibitem{KLP} F. Karsch, E. Laermann, and A. Peikert, {\it Nucl. Phys. B} 
{\bf 605}, 579 (2001).
\bibitem{GI} S. Godfrey and N. Isgur, {\it Phys. Rev. D} {\bf 32}, 189 (1985).
\bibitem{Xu2002} X.-M. Xu, {\it Nucl. Phys. A} {\bf 697}, 825 (2002).
\bibitem{SX} Z.-Y. Shen and X.-M. Xu, {\it J. Korean Phys. Soc.} {\bf 66}, 754
(2015).
\bibitem{SXW} Z.-Y. Shen, X.-M. Xu, and H. J. Weber, {\it Phys. Rev. D} 
{\bf 94}, 034030 (2016).
\bibitem{WX} T.-T. Wang and X.-M. Xu, {\it Chin. Phys. C} {\bf 43}, 024102 
(2019).
\bibitem{YXW} K. Yang, X.-M. Xu, and H. J. Weber, {\it Phys. Rev. D} {\bf 96},
114025 (2017).
\bibitem{CMS2018} CMS Collaboration, {\it Eur. Phys. J. C} {\bf 78}, 509 
(2018).
\bibitem{ATLAS2018} ATLAS Collaboration, {\it Eur. Phys. J. C} {\bf 78}, 762 
(2018).
\bibitem{JXW} S.-T. Ji, X.-M. Xu, and H. J. Weber, {\it Nucl. Phys. A} 
{\bf 966}, 224 (2017).
\end{thebibliography}
\end{document}